\documentclass[aps,A4,pra, superscriptaddress ,preprintnumbers,floatfix,footinbib,twocolumn]{revtex4}
\usepackage{amsfonts}
\usepackage{amsmath}
\usepackage{amssymb}
\usepackage{graphicx}
\usepackage[T1]{fontenc}

\begin{document}

\title{The focusing effect of cold atomic cloud with a red-detuned Gaussian
beam}
\author{Zhenglu Duan$^{\ast}$}
\affiliation{Center for Quantum Science and Technology, Key Laboratory of Photoelectronic and Telecommunication of Jiangxi province, Jiangxi Normal
University, Nanchang, 330022, China }
\author{Shuyu Zhou$^{\dagger}$ }
\affiliation{Key Laboratory for Quantum Optics, Shanghai Institute of Optics and Fine Mechanics, The Chinese Academy of
Sciences, Shanghai 201800, China} 
\author{Tao Hong}
\affiliation{Key Laboratory for Quantum Optics, Shanghai Institute of Optics and Fine Mechanics, The Chinese Academy of
Sciences, Shanghai 201800, China} 
\author{Yuzhu Wang}
\affiliation{Key Laboratory for Quantum Optics, Shanghai Institute of Optics and Fine Mechanics, The Chinese Academy of
Sciences, Shanghai 201800, China} 

\begin{abstract}
We have demonstrated an atom-optical lens, with the advantage of a small
scale and flexible adjustment of the parameters, realized by a far
red-detuned Gaussian laser beam perpendicular to the propagation direction
of the cold atomic cloud. The one-dimensional transverse focusing effect of
cold atomic clouds at the temperature order of 1 $\mu$K freely falling
through the atom-optical lens on the micron scale have been studied
theoretically and then verified experimentally. It is found that theory and
experiment are in good agreement.
\end{abstract}

\maketitle

\section{ Introduction}

An atom-optical lens is one of the fundamental atom-optical elements, which
can focus, collimate, image and transmit the atom beam or atomic cloud \cite%
{Meystre, PR240-143}. Up to now two main types of atom-optical lens, based
on magnetic or optical fields, have been developed. Atom-optical lenses
based on magnetic fields \cite{PRL67-2439, PRL87-030401,
PRA65-031601,APB2007} are advantageous for coherent atom-optics research
owing to their extremely high optical quality. However, it is difficult to
build flexible optical systems because magnetic atom-optical elements have a
large scale. In contrast, atom-optical lens based on the optical dipole
force has a small scale and is flexible to realize the combination of
atom-optical lenses \cite{PR240-143, BFAP78, JOSAB8-502}. Except for optical
dipole force, radiation-pressure force \cite{Wpzhang1,Wpzhang2,JETP43-217, JMO35-17},
near-field light \cite{JOA8-153, PRA77-013601}, and far-detuned and resonant
standing wave fields \cite{PRA60-4886} are also utilized to realize an
atom-optical lens.

Focusing of an atomic beam or cloud are one of most important applications
of atom-optical lens, which can offer high bright sources. Such sources are
desired by atom lithography\cite{lith}, atom interferometry, atomic fountain
clock\cite{clock}, atomic physics collision experiments\cite{collision},
ultra high resolution optical spectrum and quantum frequency standard. In
fact, all these applications of atom-optical lens do not require a small
scale. However, when considering loading or transmitting atomic cloud onto
micro-trap or micro-waveguide on a integrated atom chip, the atom-optical
lenses on the micron scale are expected. For a Gaussian laser beam with red
detuning, the dipole force is toward the maximum intensity region. Hence, a
focused Gaussian laser beam with red detuning can be used as an atom-optical
lens. To avoid the aberration of the atom-optical lens from the spontaneous
emission, the detuning should be large enough. Experimentally, the waist of
the focused laser beam can be focused to several microns. Consequently, a
focused red-detuned Gaussian laser beam is suitable to make the atomic beam
or cloud focus on the micron scale.

In our previous work we have experimentally observed focusing and
advancement effects when ultracold atomic clouds and BEC wave packets passed
through focused red-detuned Gaussian laser beam\cite{PRA80-033411}. In that
work the dynamics of ultracold cloud and BEC are described by the linear
(nonlinear) Schordinger equation. While this work will theoretically study
the one-dimensional focusing of cold atomic clouds passing through an
atom-optical lens with Newtonian mechanics and then experimentally verify
the theoretical prediction. The atom-optical lens suggested in this work has
the advantage of easily constructing and aligning the setup because the
laser beam is perpendicular to the propagation direction of the atomic cloud.

The remainder of the paper is organized as follows. In Sec. II we study the
focusing of the atom-optical lens induced by far red-detuned Gaussian laser
beam using particle tracing method when the atom is under gravity field. In
Sec III the experimental investigation of the focusing effects is presented
and discussed. Finally we conclude the work.

\section{Theory analysis}

We first consider an atom located at the position $(0,0,L)$ freely falling
along $z$ axes to a potential induced by a far red-detuned focused Gaussian
laser beam:%
\begin{equation}
U=-U_{0}\exp \left( -\frac{y^{2}+z^{2}}{\sigma _{0}^{2}}\right) ,  \label{V}
\end{equation}%
where interaction intensity $U_{0}=\hslash \Omega ^{2}/4\left\vert \delta
\right\vert $. Here $\Omega $ is determined by the intensity in the center
of the Gaussian beam, and $\sigma _{0}$ is the waist width of the Gaussian
beam. $\delta =\omega _{L}-\omega _{a}$ presents the detuning between the
laser frequency and the transition frequency of the atom. When the detuning
is red, the potential is negative and presents an attractive force when the
atoms passing through it. The red-detuned Gaussian laser beam can therefore
serve as a focal cylindrical atom-optical lens. To avoid the aberration of
the atom-optical lens from the spontaneous emission, the detuning $\delta $\
should be far larger than the spontaneous emission rate.

\begin{figure}[tbp]
\begin{center}
\includegraphics[width=3.3in]{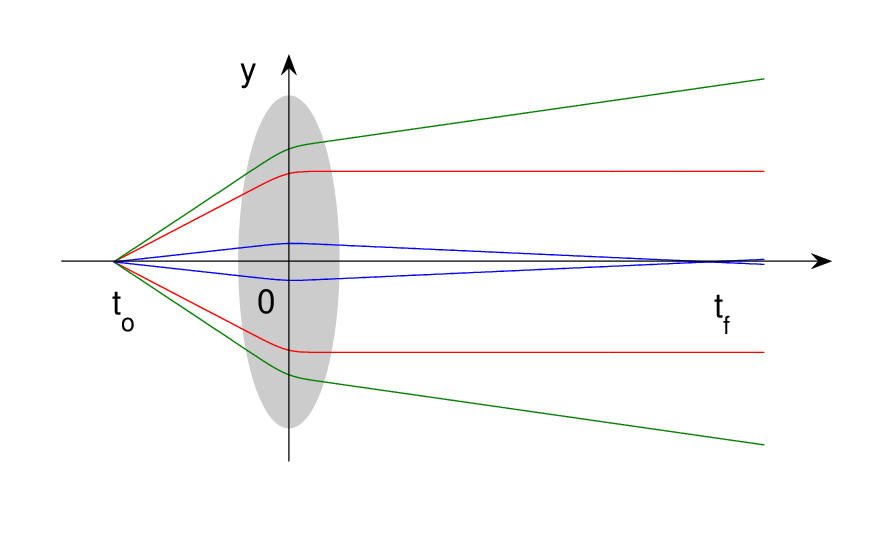}
\end{center}
\caption{{\protect\footnotesize (Color online) Space-time diagram of the
atomic trajectory. The object point is focused on imaging point by the atom
lens.}}
\label{fig1}
\end{figure}

Now we will investigate the focusing effect of the atom-optical lens by
solving the motion equation of atoms:

\begin{equation}
m\frac{d^{2}\vec{r}}{dt^{2}}=-\vec{\nabla}\left( U-mgz\right) ,  \label{r}
\end{equation}
where $m$ is the atom mass. Due to free expansion of the atomic cloud along $%
x$ direction, hereafter we will denote $y$ direction as transverse direction.

We assume that, without loss of important physics, the initial height of the
atoms is sufficiently large so that, when atoms pass through the laser beam,
their kinetic energies are far greater than the optical potential, and
therefore their velocity along $z$ direction almost remains unchanged \cite%
{moment}. Under this assumption, we obtain the transverse momentum change
along $y$ direction on the atomic center of mass can be found with the law
of impulse-momentum

\begin{eqnarray}
m\Delta v_{y} &=&-\int_{-\infty }^{\infty }\frac{\partial }{\partial y}%
U\left( t\right) dt  \notag \\
&\simeq &mv_{y0}f  \label{delta_v}
\end{eqnarray}%
\bigskip with%
\begin{equation}
f=-\frac{2\sqrt{\pi }U_{0}}{mg\sigma _{0}}\exp \left( -\frac{%
v_{y0}^{2}t_{o}^{2}}{\sigma _{0}^{2}}\right)  \label{f}
\end{equation}%
where $t_{o}=\sqrt{2L/g}$ is the falling time spent by the atom to reach the
center plane of the laser beam from its initial position and $v_{y0}$ is its
initial transverse velocity.

Using the simple geometric relationship of time-space trajectories of atoms,
as shown in figure \ref{fig1}, one finds that the relation between flying
time $t_{o}$ and the focusing time $t_{f}$ in imaging region as

\begin{equation}
t_{f}=-\frac{t_{o}}{\left( f+1\right) }  \label{tf}
\end{equation}

\begin{figure}[tbp]
\begin{center}
\includegraphics[width=3.5in]{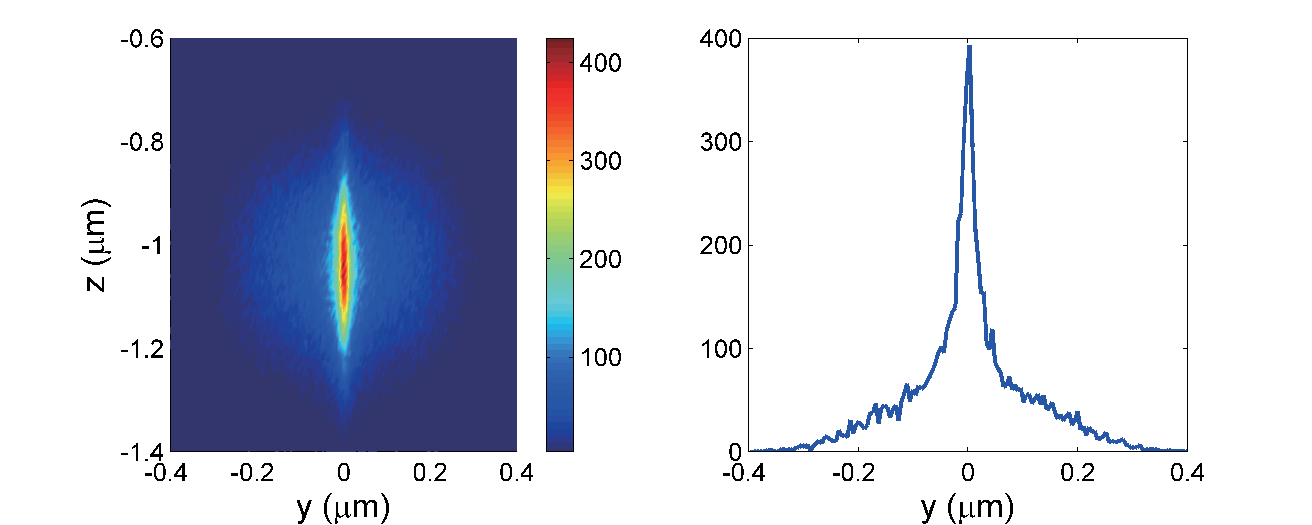}
\end{center}
\caption{{\protect\footnotesize (Color online) The simulation of focusing
atomic clouds by the Gaussian laser beam. (a) shows images of the simulated
cold atomic cloud passing through the laser beam. (b) is the corresponding
cross section of focused atomic cloud along $y$ direction at the widest part
of atomic cloud. The initial temperatures of the atomic clouds is $0.5%
\protect\mu $ k. The flying times of the atomic cloud are $t_{0}=7$ ms and $%
t=10$ ms. Other parameters are $U_{0}=3.0\times 10^{-29}$J and $\protect%
\sigma _{0}=50\protect\mu m$.}}
\label{fig2}
\end{figure}
From Eq. (\ref{tf}) one can note that, if $f<-1$, $t_{f}>0$, the atomic
trajectory will be bent to the axes of the atomic lens in the imaging
region, i.e., a real image of the atom formed by the atomic lens; if $f>-1$,
$t_{f}<0$, the atomic trajectory will be bent from the axes in imaging
region. Hence the dimensionless parameter $f$\ characterizes the the atom
deflection by the laser beam. Eq. (\ref{f}) suggests that the parameter $f$
can be easily controlled by adjusting the interaction intensity $U_{0}$ or
the laser beam waist $\sigma _{0}$, and therefore one can focus or defocus
the atomic cloud by adjusting the parameters of the laser beam.

Eq. (\ref{tf}) also shows that the time for the atomic cloud to be focused
after it passes through the laser beam is proportional to the initial flying
time, which is different from the ordinary object-image relationship of
focal atom-optical lens in the time domain \cite{PRA59-4636}. This is
because the atoms are accelerated in gravity in our case while the velocity
is uniform in other cases. Additionally, from Eqs. (\ref{tf}) and (\ref{f})
one can find that the focusing time is dependent on the initial transverse
velocity of the atom. Consequently, an atomic cloud will be focused to one
small spot, instead of one point. This leads to the spherical aberration of
the atom-optical lens.

Now we consider the case of cold atomic ensemble. Since the parameter $f$ is
dependent on the transverse velocities of the atoms, the atoms in the
ensemble may be focused or not by the atomic lens dependent on its initial
transverse velocity. Figure \ref{fig1} presents some typical atomic
trajectories with different initial transverse velocities. From figure \ref%
{fig1} one can see that only the atoms whose initial transverse velocity is
smaller than critical velocity $v_{c}$ can be focused in the imaging region
by the atomic lens. The critical velocity is determined by

\begin{equation}
f=-1
\end{equation}%
and the corresonding velocity is%
\begin{equation}
v_{c}=\frac{\sigma _{0}}{t_{0}}\sqrt{\ln \left( \frac{2\sqrt{\pi }U_{0}}{%
mg\sigma _{0}}\right) }.  \label{vc}
\end{equation}

Eq. (\ref{vc}) shows that, for real experimental parameters, the critical
velocity has a finite maximum value, in other words, only part of the atoms
can be focused. We have numerically simulated the partially focusing effect
with a cold atomic cloud passing through the laser beam. From figure \ref%
{fig2} we can see that there is a sharply focused peak on the wide unfocused
atomic cloud background. Since the focusing effect only happens along $y$
direction, the atomic cloud looks like a long cigar.

\begin{figure}[htbp]
\begin{center}
\includegraphics[width=3.2in]{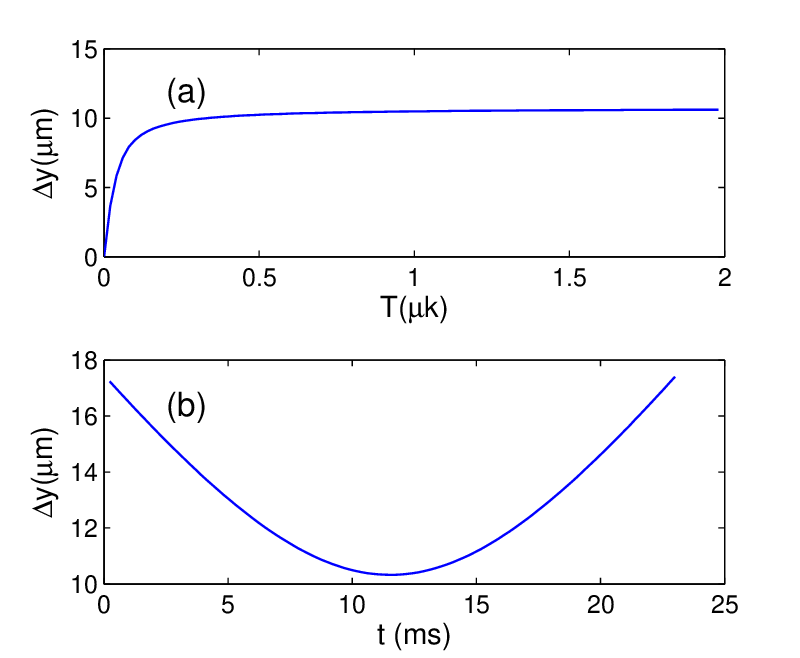}
\end{center}
\caption{{\protect\footnotesize (Color online) The transverse width of peak
of the focused atomic cloud dependent on (a) the initial temperature with $%
t=10$ ms and (b) the falling time with $T=1.0$ $\protect\mu$k. Other
parameters are $U_{0}=2.81\times 10^{-29}$J, $\protect\sigma _{0}=35\protect%
\mu m$ and $t_o = 7$ms.}}
\label{fig3}
\end{figure}

Now we calculate the transverse width of the peak of focused atomic cloud
along $y$ direction. We assume that the atomic cloud with initial
temperature $T$ is released at the position ($0,0,gt_{o}^{2}/2$), it reaches
the center of the laser beam after time $t_{o}$ and then to the image region
after time $t$. Therefore the half width of the focused atomic cloud in the
image region is

\begin{align}
\Delta y^{2}& =\left\langle v_{y}^{2}\left( t_{f}-t\right) ^{2}\right\rangle
-\left\langle v_{y}\left( t_{f}-t\right) \right\rangle  \notag \\
& =A\int_{-v_{c}}^{v_{c}}v_{y_{0}}^{2}\left( tf+t+t_{o}\right) ^{2}\exp
\left( -\frac{mv_{y_{0}}^{2}}{2k_{B}T}\right) \,\mathrm{d}v_{y_{0}}  \notag
\\
& =\frac{Af_{0}^{2}t^{2}}{2C_{1}^{3}}\left(
-2v_{c}C_{1}e^{-v_{c}^{2}C_{1}^{2}}+\sqrt{\pi }\mathrm{Erf}\left(
v_{c}C_{1}\right) \right)  \notag \\
& +\frac{2Af_{0}t\left( t+t_{o}\right) }{2C_{2}^{3}}\left(
-2v_{c}C_{2}e^{-v_{c}^{2}C_{2}^{2}}+\sqrt{\pi }\mathrm{Erf}\left(
v_{c}C_{2}\right) \right)  \label{Dy} \\
& +\frac{A\left( t+t_{o}\right) ^{2}}{2C_{3}^{3}}\left(
-2v_{c}C_{3}e^{-v_{c}^{2}C_{3}^{2}}+\sqrt{\pi }\mathrm{Erf}\left(
v_{c}C_{3}\right) \right)  \notag
\end{align}%
with
\begin{subequations}
\label{C}
\begin{align}
C_{1}& =\sqrt{\frac{2t_{o}^{2}}{\sigma _{0}^{2}}+\frac{t_{o}^{2}}{\sigma
_{a}^{2}}} \\
C_{2}& =\sqrt{\frac{t_{o}^{2}}{\sigma _{0}^{2}}+\frac{t_{o}^{2}}{\sigma
_{a}^{2}}} \\
C_{3}& =\frac{t_{o}}{\sigma _{a}}
\end{align}%
and

\end{subequations}
\begin{equation}
f_{0}=\frac{2\sqrt{\pi }U_{0}}{mg\sigma _{0}}
\end{equation}%
where the normalized constant $A=t_{o}/\left( \sigma _{a}\sqrt{\pi }\mathbf{%
Erf}\left( v_{c}t_{o}/\sigma _{a}\right) \right) $ and the size of the
atomic cloud when it reaches the laser beam center $\sigma
_{a}^{2}=2t_{o}^{2}k_{B}T/m$ with Boltzmann constant $k_{B}$.

\begin{figure*}[htp]
\begin{center}
\includegraphics[width=5in]{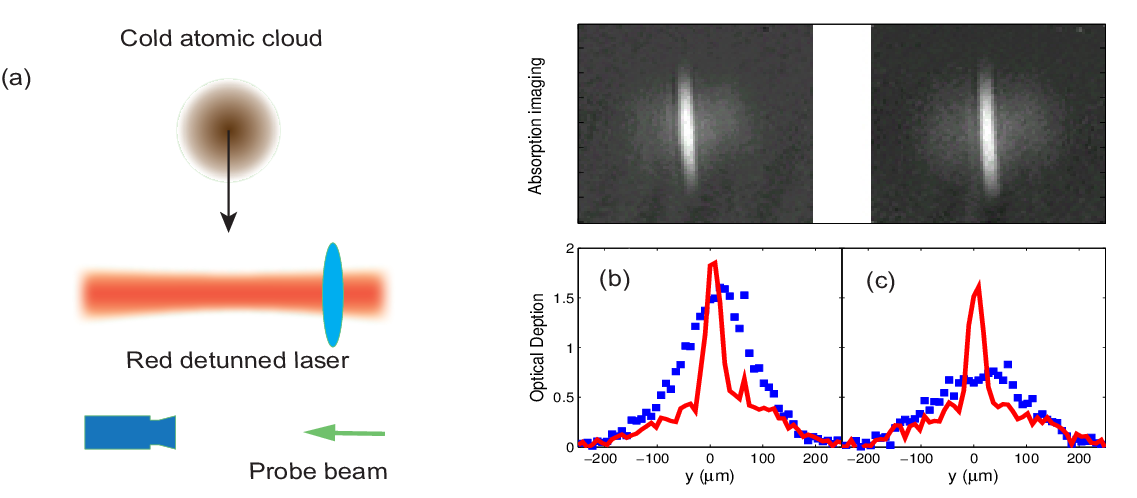}
\end{center}
\caption{{\protect\footnotesize (Color online) The focusing of atomic clouds
by the Gaussian laser beam. (a) is the schematic diagram of experimental
setup. Top panels of (b) and (c) are the absorption imaging of the focused
atomic clouds. The bottom panels are the corresponding cross section of
optical depth. The solid lines denote the the focused atomic clouds, while
the symbol $\blacksquare $ represent the reference ones. The initial
temperatures of the atomic clouds are (a) 190nk and (b) 370nk respectively.
The atomic clouds flies 7 ms before they get to the center of laser beam and
then 9 ms to the imaging region. Other parameters are $U_{0}=2.81\times
10^{-29}$J and $\protect\sigma _{0}=35\protect\mu m$.}}
\label{fig4}
\end{figure*}

We plot the transverse width of the peak of the focused atomic as functions
of initial temperature $T$ and falling time $t$, shown in figure \ref{fig3},
respectively. It can be noted that, from Fig. \ref{fig3}(a), the transverse
width is first rapid increases and then approximately approaches to a
constant with increasing the initial temperature. Mathematically, this can
be seen from Eq. \ref{Dy} that, when $T$ is very small, $\Delta y^{2}\propto
\sigma _{a}^{2}\propto T$, while $T$ is very large, $\Delta y^{2}\propto
\exp \left( -mv_{y_{0}}^{2}/\left( 2k_{B}T\right) \right) \rightarrow 1$,
i.e., $\Delta y$ is independent on $T$. Physically, when the initial
temperature of the atomic cloud is very lower, almost all the atoms are
focused by the laser beam, and in this case the transverse chromatic
aberration dominantly contributes to the transverse width of the focused
atomic cloud. When the initial temperature is very high, only the center
part of the atomic cloud overlaps the laser beam and is focused by it. The
focused atoms tends to be monochromatic, rather chromatic, with increase of
the initial temperature. In this situation, the spherical aberration of the
atomic lens dominantly contributes to transverse width of the atomic cloud
than the transverse chromatic aberration does. Therefore the transverse
width is just related with parameters of the atomic lens, instead of the
initial temperature of the atom cloud.

When fixing the initial temperature $T$ and changing the falling time $t$,
the width of the focused atomic cloud is first decreasing and then
increasing, as shown in figure \ref{fig3}(b). The minimum width of the
focused atomic cloud along $y$ direction increases with increasing the
initial temperature. Of course, if the initial temperature is so lower that
the atomic cloud is condensed, the quantum mechanics model is required to
describe the focusing effect.

\section{ Experimental result and discussion}

Our experimental setup consists of two magneto-optical traps \cite%
{PRA80-033411}. The atomic cloud is firstly captured in the Up-MOT and then
transferred into the second ultrahigh-vacuum MOT(UHV-MOT). After the atomic
number in the UHV-MOT is stable, we prepared optical molasses and then
loaded atoms into a quadrupole-Ioffe-configuration (QUIC) trap. Evaporative
cooling of the atoms was performed by rf-induced spin flips. We swept the rf
frequency from 25MHz to a value of around 1.6MHz over a period of 28s.
Atomic clouds with various temperatures from about 1$\mu $K to below the
phase transition point were obtained by setting different rf frequencies.
The cold atomic cloud then ballistically expanded after the magnetic trap
was switched off. The direction of propagation of the focused Gaussian beam
and the probing beam was parallel to the long axis of the QUIC trap.
Therefore, the cold atomic clouds symmetrically distributed in the probing
plane before traversing the Gaussian beam. We acquired the distribution of
atomic clouds from absorption images, as shown schematically in figure \ref%
{fig4}(a). The Rayleigh length was about 8.5mm. Since the Rayleigh length
was far greater than the scale of the atomic cloud while it was passing the
light beam, we could approximate that the laser beam provided a
two-dimensional Gaussian potential. The power of the focused Gaussian light
beam was between $40\sim 45$ $\mu $w with a red detuning $\delta /2\pi
=-40\sim -80$ GHz. When evaporative cooling was finished, we released the
atomic cloud from the QUIC trap and turned on the Gaussian beam. After the
atomic cloud had passed through it, we turned off the Gaussian beam and
waited for several milliseconds. All information regarding the atomic cloud
was derived from the absorption images.

\begin{figure}[h]
\begin{center}
\includegraphics[width=3.0in]{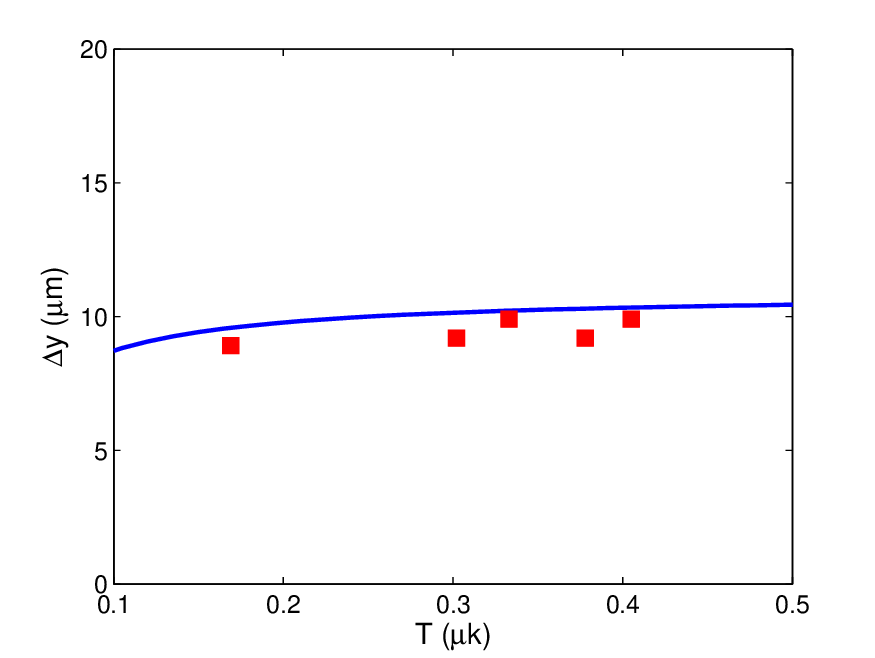}
\end{center}
\caption{{\protect\footnotesize (Color online) The transverse width of the
focused part of the atomic cloud against the initial temperature. The symbol
$\blacksquare $ gives the experimental result. The solid line is the
analytical result. Other parameters are $U_{0}=2.81\times 10^{-29}$J, $%
\protect\sigma _{0}=35\protect\mu m$, $t_{0}=7ms$, $t=9ms$.} }
\label{fig5}
\end{figure}

Figures \ref{fig4}(b) and \ref{fig4}(c) are typical experimental results
about the focusing of the atomic clouds with different initial temperatures
passing through the laser beam. From absorption images and cross section of
optical depth we can see that the atomic clouds are focused by the laser
beam compared to the reference one. As expected, the focusing just happened
along $y$ direction, while the vertical direction remained unchanged. Owing
to the finite action length of the laser beam, only the center part of the
atomic cloud is focused, which is consistent with our theoretical anlysis.
The ratio of focused parts increases with the decrease in temperature.

\begin{figure}[h]
\begin{center}
\includegraphics[width=3.0in]{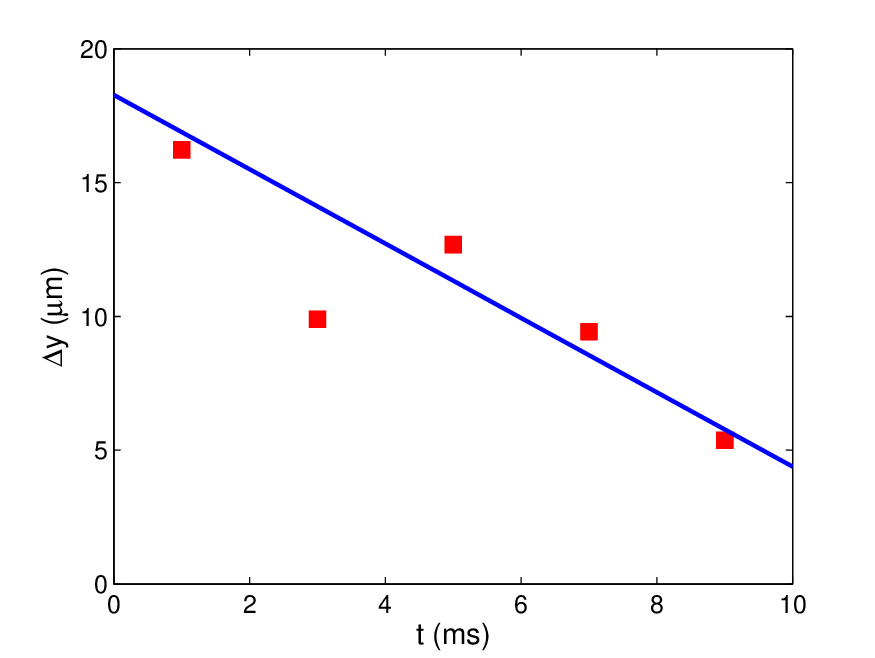}
\end{center}
\caption{{\protect\footnotesize (Color online) The transverse width of the
focused part of the atomic clouds plotted against the falling time. The
symbol $\blacksquare $ gives the experimental result. The solid line is the
analytical result. Before passing through the Gaussian beam, the atomic
clouds have already flown 7 ms. Other parameters are $U_{0}=1.26\times
10^{-29}$J, $\protect\sigma _{0}=17\protect\mu m$ and $T_{0}=71nk$.}}
\label{fig6}
\end{figure}

As the case in figure \ref{fig3}, we also measured the transverse width of
the focused part of the atomic cloud with respect to the initial temperature
of the atomic cloud and falling time after the focused atomic cloud passing
through the laser beam, respectively. The experimental results are plotted
in figures \ref{fig5} and \ref{fig6}, which are well consistent with the
analytical prediction.

\section{Conclusion}

We study the one dimensional transverse focusing effect of an atomic cloud
freely falling and passing through the atom-optical lens induced by the far
red-detuned laser beam. Based on the atom deflection in dipole conservative
potential, the relation between initial falling time and focusing time is
theoretically presented. Moreover, atom-optical lens induced by the laser
beam has the advantages of small scale and flexible adjustment. It may play
an important role in many fields such as, integrated atom optics and atomic
fountain clock. Finally we have experimentally demonstrated the focusing,
imaging of the atomic cloud passing through the Gaussian laser beam, which
are well consistent with the theoretical prediction.

This work is supported by the National Natural Science Foundation of China
under Grant Nos. 10828408, 10804115 and 10974211, National Fundamental
Research Program of China under Grant No. 2011CB921504.

\end{document}